\RequirePackage{fix-cm}
\RequirePackage[colorlinks,citecolor=blue,urlcolor=blue]{hyperref}
\documentclass{svjour3}

\smartqed  

\usepackage{graphicx}
\usepackage{multirow}
\usepackage{booktabs}
\usepackage{xcolor}
\usepackage{textcomp}
\usepackage{url}
\usepackage{enumitem}
\usepackage{tabularx}
\usepackage[normalem]{ulem}
\usepackage{empheq}
\usepackage{tipa}
\usepackage{natbib}

\usepackage{tcolorbox}
\tcbuselibrary{breakable}
\tcbset{
  width=0.85\textwidth,
  halign=justify,
  center,
  breakable
}

\usepackage{tikz,xcolor,hyperref}

\definecolor{lime}{HTML}{A6CE39}
\DeclareRobustCommand{\orcidicon}{%
    \begin{tikzpicture}
    \draw[lime, fill=lime] (0,0) 
    circle [radius=0.16] 
    node[white] {{\fontfamily{qag}\selectfont \tiny ID}};
    \draw[white, fill=white] (-0.0625,0.095) 
    circle [radius=0.007];
    \end{tikzpicture}
    \hspace{-2mm}
}

\begin{document}
  
\title{Which Combination of Test Metrics Can Predict Success of a Software Project? A Case Study in a Year-Long Project Course}
\titlerunning{Which Combination of Test Metrics Can Predict Success of a Software Project?}

\author{Marina Filipovic\href{https://orcid.org/0000-0001-5641-506X}{\orcidicon} \and Fabian Gilson \href{https://orcid.org/0000-0002-1465-3315}{\orcidicon}}

\authorrunning{Marina Filipovic \and Fabian Gilson}

\institute{ M. Filipovic \at 
    Computer Science and Software Engineering \\
    Te Whare Wānanga o Waitaha | University of Canterbury\\
    Christchurch, New Zealand\\
    \email{marina.filipovic@canterbury.ac.nz}
    \and
    F. Gilson (corresponding author)\at
    Computer Science and Software Engineering \\
    Te Whare Wānanga o Waitaha | University of Canterbury\\
    Christchurch, New Zealand\\
    \email{fabian.gilson@canterbury.ac.nz}
}

\date{Received: date / Accepted: date}

\maketitle

\begin{abstract}
\textbf{Context:} 
Testing plays an important role in securing the success of a software development project. Various techniques of automated testing have been developed, including automated acceptance testing which represent the customer's expectations where requirements can be translated into automated tests seamlessly. Prior studies have demonstrated beneficial effects of applying acceptance testing within a Behavioural-Driven Development method.

\textbf{Objectives:} 
In this research, we investigate whether we can quantify the effects various types of testing have on functional suitability, i.e. the software conformance to users' functional expectations. We explore which combination of software testing (automated and manual, including acceptance testing) should be applied to ensure the expected functional requirements are met, as well as whether the lack of testing during a development iteration  causes a significant increase of effort spent fixing the project later on. 

\textbf{Method:}
To answer those questions, we collected and analysed data from a year-long software engineering project course. Collected data per sprint included delivered story points, testing coverage metrics, and testing effort as per students work-logs. We combined manual observations and statistical methods, namely Linear Mixed-Effects Modelling, to evaluate the effects of coverage metrics as well as time effort on passed stories over 5 Scrum sprints. 

\textbf{Results:}
The results suggest that a combination of a high code coverage for all of automated unit, acceptance, and manual testing has a significant impact on functional suitability. Similarly, but to a lower extent, front-end unit testing and manual testing can predict the success of a software delivery when taken independently. However, students time work-logs do not show statistically significant relationship between the time efforts and neither of the number of user stories delivered, nor the time spent fixing their software product in the following sprint. We observed a close-to-significant effect between low back-end testing and deferral (i.e. postponement) of user stories.

\textbf{Conclusions:}
Based on the collected data, automated acceptance testing alone does not have an influence on the product's functional suitability by itself, even though its purpose is to accurately and reliably translate user's expectations into tests, as argued in the literature. Therefore, we believe that our results can be explained by the students' lack of experience with advanced automated testing techniques, and such skills are complicated to develop. We also argue that manual acceptance testing remains extremely valuable to ensure a high functional suitability, therefore the full testing pyramid must be considered for successful product delivery.


\end{abstract}

\section{Introduction}
\label{sec:introduction}

Agile software development favours a lighter upfront requirement analysis with shorter delivery cycles~\cite{Beck2000}. Among other practices embraced by the agile movement, continuous integration has improved release processes \textit{``thanks specifically to a sharper focus on metrics and process standardization''}, including regular and automated acceptance testing (referred to as \textit{``client scenarios''})~\cite{Roche2013}. 
Developers are expected to contribute to quality assurance processes by testing the code they write as they develop new features. Within that method framework, automation is a crucial factor, including for (continuous) testing purposes~\cite{Lwakatare2015}. Therefore, technologies have emerged to test both units of code (e.g., JUnit, Jest) and delivered features as requested by the user, i.e. user acceptance testing (e.g., Cucumber).

However, benefits of continuous integration can be lost when a development process exhibits some anti-patterns such as the lack of automated testing, including in production-like environments~\cite{Zampetti2020}, which create software products of low quality. Code coverage metrics are a common way to ensure a software behaves as expected where batteries of automated tests are checking all possible execution paths of the software~\cite{Bertolino2007}. However, high coverage metrics are not always linked to good tests~\cite{Inozemtseva2014}, as confirmed by Bai et al.'s survey with students who considered code coverage as the most important metric, but the same students tended to test only blue sky scenarios~\cite{Bai2021}. Therefore, in this work, we go beyond by analysing the testing effort from multiple viewpoints, i.e. unit, automated acceptance and manual testing on the one hand, and the overall testing effort in terms of time spent on testing, or fixing code on the other hand.

Our focus in this research is to empirically investigate the relationship between the testing behaviour of multiple teams of junior developers and the external quality of the software produced during a year-long software engineering project. In other words, we investigate \textbf{to what extend a combination of metrics related to automated unit testing, automated user acceptance testing, manual testing and testing effort can predict the acceptance of user stories during sprint reviews}. Therefore, for this work, we will refer to \textit{external quality} as the \textit{Functional Suitability} as defined in IEC/ISO 25010 standard for software quality~\cite{ISO25010} where functional requirements translate into acceptance criteria attached to user stories. 

The context of this study is a year-long software engineering project at the latest stages of undergrad studies. We observed 8 teams of 4 to 8 developers over 6 sprints, i.e. Scrum iteration cycles, where students apply agile software development principles including automated testing, continuous integration, test- and behaviour-driven development. We collected code coverage metrics calculated by testing frameworks (i.e. JUnit, Jest, within SonarQube reports) together with manually calculated coverage of acceptance criteria attached to user stories (from Cucumber automated acceptance tests), as well as manual test scripts (in spreadsheets or wikis). We also analysed the teams' testing effort in terms of hours logged in Jira with special \textit{tags} passed in their work-logs (e.g., \texttt{\#test}, \texttt{\#testmanual}, \texttt{\#fix}). In details, for each sprint:

\begin{enumerate}
\item we took snapshots of the code coverage of the produced software in terms of unit testing for each sprint deliverable;
\item we mapped automated and manual acceptance testing scripts and manual test records to each acceptance criteria of the user stories that were planned for each delivery;
\item we evaluated the software products with a manual assessment of delivered user stories after each development iteration, as a product owner would do during a Scrum sprint review;
\item we analysed the teams' testing effort in terms of hours logged as \texttt{\#test} and \texttt{\#testmanual}, in relation to the teams' effort to implement features and fix bugs, logged as \texttt{\#implement} and \texttt{\#fix} respectively, as retrieved from the students' work-logs in Jira and Gitlab repositories.
\end{enumerate}

The remainder of the paper is organised as follows. We discuss the related work in Section~\ref{sec:rw}. We introduce the study setting and the method followed in Section~\ref{sec:method}. We present our results in Section~\ref{sec:results} and discuss these results in Section~\ref{sec:discussion} together with their implications to educators and practitioners as well as the limitations and threats to validity. We introduce the next steps and conclude in Section~\ref{sec:conclusion}.

\section{Related Work}
\label{sec:rw}

\subsection{Secondary Studies on Testing Practices}

Some secondary studies have investigated how some testing techniques may affect the external quality of software. 
Juristo et al.~\cite{Juristo2004} then Vassalo et al.~\cite{Gonzalez2014} classified testing techniques and evaluation metrics used, but did not address their effects on external quality in their literature reviews.
Bissi et al. conducted the most recent systematic review on the effects of test-driven development on produced software, including on external quality~\cite{Bissi2016}. From that review, we discuss all relevant works to ours in the Section~\ref{sec:rw-tdd-bdd} below, i.e. \cite{Maximilien2003, Yenduri2006, Bhat2006, Slyngstad2008, Vu2009, Pancur2011}.
Abushama et al. conducted a systematic literature review on the effects of test- and behaviour-driven design~\cite{Abushama2021}, from which one work is relevant to ours~\cite{Scandaroli2019}, also discussed below.
Van Heugten Breurkes et al. conducted a systematic literature review on the overlap of automated unit and acceptance testing in industry context~\cite{vanHeugten2022}. The authors concluded that a test-first mindset can be beneficial without being able to demonstrate any further correlation.
Binamungu and Maro report on a mapping study about Behaviour-Driven Development (BDD) where they investigated the common themes and focus of prior studies where BDD has been applied~\cite{Binamungu2023}. We discuss hereafter the relevant papers to our focus, i.e.~\cite{Dookhun2019, Rocha2021}.
Farooq et al. conducted a systematic literature review of uses and applications of BDD where they concluded that BDD can address communication gaps between business stakeholders and developers and therefore improve the quality of software development processes as well the quality of the final product~\cite{Farooq2023}. We discuss hereunder relevant papers, i.e.~\cite{Carrera2014,Scandaroli2019,Nascimento2020}.

\subsection{Test- and Behaviour-Driven Development}
\label{sec:rw-tdd-bdd}

The literature on the effects of test-driven development (TDD) and, to a lower extent behaviour-driven development (BDD) is luxurious. In the following, we only consider case studies that have been conducted over an extended period of time (i.e. at least multiple weeks) in team-based environments, since this is our focus in this work.

\subsubsection{In academia}

Yenduri and Perkins reported a positive effect of TDD on the numbers of \textit{``faults''} in the code written by students over multiple months~\cite{Yenduri2006}. But no details are given on the complexity of the code written, how representative it is of the real world, or whether the produced code matched the stakeholders' functional expectations.
Vu et al. performed a comparative case study in a capstone project with the industry~\cite{Vu2009}. However, the small sample size, the difference in the technology stack, and high mortality rate prevented the authors to draw any insightful conclusions.
Huang and Holcombe ran a project-based course over 12 weeks where students gather the requirements and develop a software for an external client~\cite{Huang2009}. The work is then evaluated following qualitative criteria only, but not extensively regarding its functional suitability.
Pančur and Ciglarič investigated the effects of TDD on the external quality of programming assignments with automated acceptance tests given to students in advance~\cite{Pancur2011}. In this work, we investigate the evolution of a larger scale project over time.
Lopes de Sousa et al. introduced BDD into a Scrum-based software development of an academic tool where they observed benefits in terms of accurate translations between user stories and functional requirements~\cite{LopesdeSouza2017}.
Nascimento et al. investigated the impact of BDD within a mobile computing course~\cite{Nascimento2020}. Their analysis relied on semi-structured interviews where we investigate quantifiable evidence of functional suitability of the delivered products.
Baldassarre et al. studied the ``retainment'' of TDD (\textit{aka} ability of students to retain TDD practices in their work) by a student cohort over a six month period, without considering the functional suitability~\cite{Baldassarre2021}.
Santos et al. conducted a meta analysis of twelve experiments conducted either in academia or industry to investigate, among others, the external quality of software produced under TDD practices~\cite{Santos2021}. Unlike other primary and secondary studies, TDD had a detrimental effect, however their baseline was not \textit{``pure waterfall''}, but a test-last approach in an iterative process made of short cycles. 
Rocha et al. report on their experience with students applying TDD and BDD in a software project course where they observed better grades and faster delivery~\cite{Rocha2021}. However, the authors do not clearly specify whether the grades were directly linked to the external quality of the delivered software products.

\paragraph{Our contribution}

Our work complements some of the aforementioned experiments by examining other testing methods than TDD solely, i.e. automated acceptance and manual testing too. Furthermore, our contribution is the first to investigate the link between testing metrics (i.e. unit test coverage, implementation / testing effort, and coverage of acceptance criteria associated to user stories) and the external quality (i.e. functional correctness) of delivered products.

\subsubsection{In the industry}

Bhat and Nagappan reported on multiple case studies where TDD was applied at Microsoft, resulting in a defect rate being halved compared to similar projects using a test-last approach~\cite{Bhat2006}.
Damm and Lundberg report on two projects where TDD was introduced for component-level testing and resulted in marginal decrease of defects being discovered at a later testing stage~\cite{Damm2006}.
Kobayashi et al. introduced extreme programming practices into two software development projects and identified a positive effect on the code quality from interviews with developers~\cite{Kobayashi2006}.
Maximilien et al., then Sanchez et al. reported on the shift to test-driven development (TDD) at IBM leading to a significant decrease in defect rates over multiple projects~\cite{Maximilien2003,Sanchez2007}.
Slyngstad et al. discussed the shift to TDD after a few iterations of a development project with a significant decrease in defects rate~\cite{Slyngstad2008}.
Similarly to Santos et al.~\cite{Santos2021}, Siniaalto and Abrahamsson identified a decreasing code quality when applying TDD in a comparative case study~\cite{Siniaalto2008}.
Bannerman and Martin observed that a \textit{``test-with-development''} approach decrease the code complexity on open source projects~\cite{Bannerman2011}.
Carrera et al. proposed the BDD-inspired BEAST method for multi-agent systems that was evaluated while developing a fault diagnosis tool for fibre networks~\cite{Carrera2014}. However the authors focused on test effort in terms of produced test code, not functional suitability for stakeholders.
Scandaroli et al. report on the application of behaviour-driven design alongside TDD in two industry projects~\cite{Scandaroli2019}. They mention some benefits qualitatively (e.g., better alignment with business, higher acceptance of delivered features), but do not give evidence of such benefits in their experience report.
Dookhun and Nagowah report an improvement in the external quality of software when BDD is used in conjunction to TDD, at the cost of a decrease in productivity and internal quality~\cite{Dookhun2019}.
Papis et al. conducted an experiment where interns rotated between a no-unit-test, test-last or test-first methods to develop features in one-week iterations~\cite{Papis2022}. The authors identified a positive effect on the number of bugs introduced under test-first as well as for the code quality (assessed by pre-written acceptance tests).

\paragraph{Our contribution}

Despite being carried within an academic context, compared to all of the above work, our contribution is the first to directly observe software development projects over an extended period of time where subjects are expected to write both the code together with automated acceptance and manual tests. Furthermore, the size and complexity of the software written during our project course is typically in the range of, or bigger than the codebase used in most of the case studies mentioned above.

\subsection{Testing Effort And External Quality}

Some works have been proposed to analyse programming patterns and incentivise or give feedback to students on their programming behaviour, but not so much on their testing effort, as identified by Keuning et al.'s literature review~\cite{Keuning2018}. From the list of tools, only a few are relevant to our work, that we discuss here. 

Edwards et al. proposed Web-CAT, recently augmented with a gamification module, to encourage students to write more tests~\cite{Edwards2003,Buffardi2014,Edwards2020}. However, differences in external quality was marginal after the introduction of Web-CAT  and the gamification module still requires further validation. 
Helminen et al. trace programming and testing behaviours of students in a web-based course~\cite{Helminen2013}. But the testing effort is observed from console-based interactions, not from structured automated testing.
Kazerouni et al. defined a series of metrics to estimate the testing effort in students assignments and how these metrics can be used to predict the external quality of the software produced~\cite{Kazerouni2017,Kazerouni2019}. Similarly to many works discussed in Section~\ref{sec:rw-tdd-bdd}, the authors observed a correlation between the external quality and an incremental testing method. However, the authors argued for a similar experiment to be conducted in a longer, larger project to validate their findings, which is our objective here.

\paragraph{Our contribution} 
Unlike those purpose-built systems where students receive frequent \textit{``nudges''} on their testing behaviour, we replicate the industry by limiting these nudges to sprint reviews (i.e. passed stories), as well as feedback on testing effort in relation to the product's external quality. We are also interested in understanding the overall testing dynamics in a team-based environment, aspect that has not been analysed in previous works.

\subsection{Continuous Integration and Testing}

The context of this study is a project course with a strong focus on continuous integration, including the implementation of automated testing. Usage by and perceptions of developers and stakeholders of continuous integration (CI) and more particularly, the role of automated testing in rapid or continuous software deployment (CD) has been researched extensively.
Ståhl and Bosch conducted a literature review on CI practices in the industry where they clustered related aspects of the usage of CI processes (e.g., scope, fault handling, build frequency)~\cite{Stahl2014}.
Gmeiner et al. report on a 6 year long study of implementing a CI pipeline in an Austrian company with additional manual acceptance testing~\cite{Gmeiner2015}. 
Vassalo et al. surveyed a financial industry to understand their CI/CD practices, including how they were testing their software~\cite{Vassalo2016}. 
Teixera et al. conducted a systematic literature review on how DevOps can be characterised in terms of ``capabilities'' (e.g., testing, feedback loop) and ``areas'' (e.g., culture, process)~\cite{Teixeira2020}.

\paragraph{Our contribution} 
None of these works investigated the link between (unit and acceptance) testing methods or metrics, and the acceptance of delivered features by the user or product owner within a CI-enabled project, including how junior developers (i.e. students) apply automated testing on the long run.

\section{Method}
\label{sec:method}

\subsection{Research Questions}
\label{sec:rq}

Our objective is to explore whether we can predict the success of a product delivery in a Scrum-based software development project based on the teams' testing behaviour and the number of accepted stories by the product owner. Specifically, our goal can be refined in the following two research questions:

\begin{description}
\item[RQ1] How can we combine multiple test coverage metrics to predict the functional suitability~\cite{ISO25010} of a software product in a Scrum-based project?
\item[RQ2] How does the developers' testing effort relate to the remedial effort the team of junior developers will put into fixing a software product during next Scrum sprint?
\end{description}

To answer \textbf{RQ1}, we combine the following metrics into a \textit{testing effort} indicator and compare that metric to the number of successfully delivered stories (in story points) after each sprint:

\begin{itemize}
    \item code coverage from automated unit testing, i.e. functional correctness;
    \item coverage of acceptance criteria by automated acceptance testing (in \textit{Cucumber}), i.e. functional completeness;
    \item coverage of acceptance criteria by manual end-to-end system testing (retrieved from test records and spreadsheets), i.e. functional completeness;
    \item proportion of accepted user stories by the product owner, including their functional suitability;
    \item the time spent by developers writing automated tests or running manual tests.
\end{itemize}

To answer \textbf{RQ2}, we compare the \textit{testing effort} indicator calculated above and compare it to:
\begin{itemize}
    \item the proportion of hours a team of junior developers spent testing their product, i.e. writing automated tests or running manual tests;
    \item the proportion of hours a team of junior developers spent fixing bugs.
\end{itemize}

\subsection{Study Design}

\subsubsection{Participants}

This study has been carried on a whole year project course in the third year of Software Engineering degree at the University of Canterbury. At this point in the curriculum students already have a background in object-oriented software design in both Python and Java. The course has the following prerequisites: introduction to object-oriented design, programming in Java, and relational databases. As a co-requisite, students take a software engineering course that covers Scrum, automated acceptance testing, continuous integration, and design patterns. At the start of the study, 80 students were fully enrolled in the course at the beginning of the year, with 70 completing the course from which 7 were identified as women, and 63 as men. We obtained appropriate approval from students to use their data for research purposes after anonymising them.

\subsubsection{Timeframe and Support}

The project ran for the full academic year (i.e. two semesters) over 6 sprints (i.e. iterations), each sprint being approximately 3 academic weeks long. All teams started with 7 to 8 students working on the same project at first, from a small template containing the initial technology stack. Then all teams diverged to create distinguishable products from mid-year onward. The typical full code base ranged from 20 to 30K LOC. Students were expected to regularly log their work to achieve around 300 hours of work by the end of the year (University regulations). Some students dropped the course after a few sprints, typically when they were falling behind the course expectations in terms of regular engagement. This has affected some teams more than others and created differences between teams in terms of number of students at the end of the year.

Each scrum team was assigned a scrum master being a tutor, i.e. a student who achieved well in the course previously. A course lecturer (the product owner) composed the product backlog and outlined acceptance criteria (AC) for each story. At the beginning of each sprint, teams discussed their sprint commitment with the product owner, i.e. the number of stories they commit to deliver by the end of the sprint. Additional teaching staff were overlooking the technology stack used by students, training and supervising the Scrum masters, and marking the software products (documentation, test and code quality, and functional suitability of the product).

Specific workshops were offered along the year, including the acceptance testing strategies workshop, ran at the beginning of the project (sprint 2). During this workshop students learnt how to write automated acceptance tests using \textit{Cucumber}\footnote{See \url{https://cucumber.io/}}. Both automated unit and manual acceptance testing were techniques taught and practised in previous courses that are prerequisites and co-requisites for this course. 

\subsubsection{Software Development Process}
\label{sec:dev-process}

The students were expected to follow the Scrum framework while working on the project course. Each sprint, students have gone through the following Scrum events:   
\begin{description}
    \item[Backlog grooming:] Teams go through a number of stories in the backlog, understand what work is required to complete them, update the acceptance criteria if needed.
    \item [Planning:] Teams agree on the stories they will work on in the current sprint with the product owner. Teams are then responsible to compose their sprint backlog by choosing stories for the sprint. Teams also have the opportunity to discuss their sprint commitment during a sprint and defer some stories, i.e. remove them from the sprint backlog, also called deferred stories.
    \item [Sprint:] Teams work on the next increment of the product. Each student is required to participate to all aspects of product development, e.g., features development, testing, documentation, fixing bugs, refactoring.
    \item [Stand-up:] Teams run status meetings (usually 15 min long), most of them supervised by Scrum masters or another teaching team member.
    \item [Sprint review:] Teams review each others' products against the acceptance criteria of each committed story (excluding the deferred stories). An additional product review is conducted by the teaching team.
    \item [Sprint retrospective:] Teams reflect on the sprint that has just finished and decide on action items to improve their process for next sprint.
\end{description}

\subsubsection{Delivery}

Each delivery is typically composed of a software product deployed on a virtual machine. The deployment follows a continuous integration (CI) pipeline triggered from a \textit{GitLab}\footnote{See \url{https://about.gitlab.com/}} repository hosted locally at the University. The continuous integration pipeline also contains dedicated automated testing stages, and is connected to a local \textit{SonarQube}\footnote{See \url{https://www.sonarsource.com/products/sonarqube/}} instance for students to keep track of their code quality metrics, including test coverage. Alongside the code, students have to keep a trace of their design decisions, manual testing logs, user interface mockups, user guide, and any additional documentation needed to run the project locally, all of these being part of the assessed material for each delivery (i.e. sprints).

Each delivery is followed by a comprehensive assessment by the teaching team including individual performance and product feedback. The product feedback reports on the functional suitability of the product, the quality of the code, the quality and amount of testing, and the quality of the documentation. Students' performance is evaluated in terms of their ability to engage with the Scrum process and agile principles.

\subsection{Data Sources}

We use multiple sources for data collection, including first-degree \textit{work diaries}~\cite{Lethbridge2005}, automatically calculated coverage metrics\footnote{See \url{https://www.sonarqube.org/}}, and teams test scripts. In this project course, students log their time spent on the project into a project management tool. We used Jira\footnote{See \url{https://www.atlassian.com/software/jira}} where teams created their sprint backlogs and recorded their work-logs. Work-logs are composed of a short summary of the work done and the time spent on that task, i.e. Jira ticket. Students are encouraged to log their time in a granular way, so a task will have multiple work-logs. Additionally, students used prescribed tags to describe the work completed, close to Conventional Commits tags\footnote{See \url{https://www.conventionalcommits.org/}}, e.g., \texttt{\#implement} for implementation of features, \texttt{\#test} for writing automated tests, \texttt{\#testmanual} for writing or executing manual tests, and \texttt{\#fix} for bug fixing. One work-log can have multiple tags where appropriate. For example, when students use test-driven development, they will typically tag a work log with both \texttt{\#test} and \texttt{\#implement} tags. In the following, we detail how we obtained the data from both automated sources and manual recollection. For all manual recollection, one researcher carried the extraction, and a second researcher proofread the extraction and aggregation of data.

\subsubsection{Code coverage metrics} 

This is obtained by building sprint deliverable code. We access unit testing code coverage from JaCoCo\footnote{See \url{https://www.jacoco.org/jacoco/}} reports uploaded to SonarQube when a deliverable is built by the CI pipeline. Code coverage is gathered from JUnit\footnote{See \url{https://junit.org/}} (Java) and Jest\footnote{See \url{https://jestjs.io/}} (JavaScript) unit tests. For the purpose of this study we compiled the code coverage data for the delivered software products at the end of each sprint.

\subsubsection{Coverage of acceptance criteria from automated acceptance tests} 

Students are required to write automated acceptance tests in \textit{Gherkin}, i.e. \textit{Given-When-Then}, syntax using Cucumber scenarios. Metrics about the number of scenarios are collected in the aforementioned JaCoCo reports. However, there is no explicit relation between the number of scenarios and the coverage of an acceptance criteria by a scenario, so we manually mapped scenarios to acceptance criteria attached to user stories. The acceptance criterion was considered fully covered if there was at least one acceptance test written for it.

\subsubsection{Coverage of acceptance criteria from manual tests} 

To determine manual test coverage, we looked into the manual testing records teams made available on their GitLab wiki. Similarly to above, manual testing coverage was determined by mapping manual tests to acceptance criteria. The acceptance criterion was considered fully covered if there was at least one manual test written for it.

\subsubsection{Overall coverage of acceptance criteria} 

To determine the overall acceptance test coverage, we have combined both the metrics from automated acceptance testing (using Cucumber) and manual testing by determining how many ACs were covered using either or both techniques combined. For example, if there are 10 ACs for a story (say AC1-10), AC1-5 are covered by automated tests, AC4-7 covered by manual tests, then 70\% of all ACs are considered covered. This approach gives us a global indicator of the coverage of ACs for each product delivery, regardless of the technique, i.e. automated or manual.

\subsubsection{Assessment of functional suitability} 
\label{sec:func-suitability}

To determine how many stories passed or failed at the end of each sprint, markers from the teaching team reviewed the products based on the stories teams committed to at the beginning of the sprint. Markers were randomised between teams and sprints. During a sprint, teams had the ability to re-negotiate their initial sprint commitment with the product owner, and therefore, defer stories. Some of these deferred stories would eventually be picked up during a later sprints, often with a lower estimation, i.e. smaller number of story points, reflecting the estimated additional effort to pass that story. The passing rate of stories is calculated from the final commitment, i.e. without counting the deferred stories as failed stories.

\subsubsection{Time effort}
\label{sec:time-effort} 
We extracted the number of hours teams have spent on various activities from their Jira work-logs. Each team member had to classify their work-logs with specific tags, similar to Conventional Commits. We sum up hours logged for (1) implementation of features, i.e. \texttt{\#implement} work-logs tags; (2) writing automated unit and acceptance tests, i.e. \texttt{\#test}; (3) writing or carrying manual (end-to-end) testing, i.e.\texttt{\#testmanual}; and (4) bug fixing, i.e. \texttt{\#fix}. Students could use other specific tags for other types of activities, such as \texttt{\#document} to record design decisions, or \texttt{\#chore} to work on their CI pipeline, but they are not relevant for this research. Also, students could tag their work-logs with more than one tag, where suitable, e.g., \texttt{\#implement} and \texttt{\#test} when applying a test-driven development method, or \texttt{\#implement} and \texttt{\#fix} when fixing and expanding on a failed story. All team members are required to participate in all aspects of software development, so producing work-logs with a variety of tags to demonstrate their fulfilment of the course objectives.

\subsection{Metrics}

From the data sources above, we compute the following metrics for testing, functional suitability, and time effort.

\subsubsection{Testing}

\paragraph{Front-End Unit Test (\textbf{FEUT})} Unit test coverage of front-end code (i.e. user interface) of the project, in percentage of covered lines of code. The data was obtained from \textit{SonarQube}.

\paragraph{Back-End Unit Test coverage (\textbf{BEUT})} Unit test coverage of back-end code (i.e. business logic layers) of the project, in percentage of lines of code. The data was obtained from \textit{SonarQube}.

\paragraph{Automated Acceptance Test coverage (\textbf{AAT})} Automated test coverage of acceptance criteria (\textit{Cucumber} scenarios), in percentage of covered acceptance criteria. The data was aggregated from manual inspection of automated scenarios, i.e. \textit{Gherkin} features, mapped to acceptance criteria for all stories.

\paragraph{Manual Acceptance Test coverage (\textbf{MAT})} Manual tests coverage of acceptance criteria, in percentage of covered acceptance criteria. The data was aggregated from manual inspection of manual test scripts mapped to acceptance criteria for all stories.

\paragraph{Overall Acceptance Test coverage (\textbf{OAT})} Combined coverage of acceptance criteria from automated and manual acceptance tests, in percentage of acceptance criteria for all stories.

\subsubsection{Functional Suitability}

\paragraph{Passed Story Points (\textbf{PSP})} Percentage of story points passed in the sprint review. Data is based on sprint marking information obtained from manual assessment by the marker.

\paragraph{Deferred Story Points (\textbf{DSP})} Percentage of story points deferred to a later sprint. Data is gathered from teams lowering their commitment during a sprint.

\paragraph{Re-committed Story Points (\textbf{RSP})} Percentage of story points a team took on in the current sprint which were made out of either deferred or failed stories in the previous sprint and have been re-committed in the current sprint. That percentage is relative to the committed story points, excluding deferred stories, e.g., a team committed to 120 points, including 30 points belonging to re-committed stories, but deferred 20 points, then $RSP = 30 \div (120-20) = 30\%$. Points may differ from the sum of deferred and failed stories from previous sprint as not all failed/deferred stories were re-committed to, and/or their points estimates were changed from one sprint to the following.

\subsubsection{Time Effort}
\label{sec:metric-time-effort}

Sprints were not lasting for the same number of days and teams were of different sizes, so we use proportional time for time effort metrics.

\paragraph{Automated Testing Effort (\textbf{ATE})} Relative number of hours spent writing test code for unit and acceptance tests compared to the total number of hours logged by the whole team during one sprint. We do not distinguish time effort by type of automated testing as teams often go back and forth between unit and acceptance testing when validating features. This data is supplied by team members in their work-logs with \texttt{\#test} tags.

\paragraph{Manual Testing Effort (\textbf{MTE})} Relative number of hours spent writing and running manual tests by the whole team. We do not distinguish between writing and running manual tests as this can be conducted in parallel, from our observations of students. This data is supplied by team members in their work-logs with \texttt{\#manual-test} tags.

\paragraph{Fixing Effort (\textbf{FE})} Relative number of hours spent fixing feature (or test) code by the whole team. This usually happen when teams re-committed to failed stories and to a lower extent when teams discover issues during a sprint. This data is supplied by team members in their work-logs with \texttt{\#fix} tags.

\paragraph{Implementation Effort (\textbf{IE})} Relative number of hours spent writing code for feature development by the whole team. As noted in Section~\ref{sec:time-effort}, these hours are sometimes combined with other tags, including \texttt{\#test} or \texttt{\#fix} tags. This data is supplied by team members in their work-logs with \texttt{\#implement} tags.

\section{Data Analysis} 
\label{sec:results}

At the beginning of the project, teams start in sub-teams for on-boarding and engagement reasons and merge in their full teams from sprint 2, hence we collect data from the second sprint onward. We had 10 teams in total but two teams recorded incomplete data about their acceptance and manual testing, so we removed these two teams from the analysis as we were not able to calculate their testing metrics accurately.  We first look at the evolution of the testing metrics and functional suitability for all teams (RQ1), then, we report on students' work-logs about their time effort (RQ1, RQ2), and finally we conduct statistical modelling (RQ1, RQ2).

\subsection{Evolution of Testing Metrics}

At the end of each sprint, we aggregated and calculated testing metrics, including average values over all sprints, for each of the eight teams over five sprints, as shown in Table~\ref{tbl:team-test-metrics}.

\begin{table}[!ht]
\caption{Test (FEUT, BEUT, AAT, MAT, OAT) and stories-related (SP, DP, RP) metrics.}
\label{tbl:team-test-metrics}
\begin{tabular}{c r | r r r r r | r r r}
Team & S\# & FEUT & BEUT & AAT & MAT & OAT & PSP & DSP & RSP \\\hline
\multirow{6}{*}{A} & 2 & 19.70\% & 72.30\% & 5.00\% & 0.00\% & 5.00\% & 0.00\% & 0.00\% & 0.00\% \\
                   & 3 & 36.30\% & 70.30\% & 2.86\% & 0.00\% & 2.86\% & 8.68\% & 0.00\% & 23.08\% \\
                   & 4 & 47.30\% & 69.20\% & 41.67\% & 31.67\% & 62.22\% & 40.26\% & 21.43\% & 0.00\% \\
                   & 5 & 54.50\% & 76.20\% & 59.26\% & 40.00\% & 69.71\% & 80.95\% & 0.00\% & 11.90\% \\
                   & 6 & 61.30\% & 80.70\% & 43.23\% & 40.48\% & 65.85\% & 88.06\% & 0.00\% & 0.00\% \\\cmidrule{2-10}
                   & \textit{avg} & 43.82\% & 73.74\% & 30.40\% & 22.43\% & 41.13\% & 43.59\% & 4.29\% & 7.00\% \\\hline
\multirow{6}{*}{B} & 2 & 0.00\% & 35.10\% & 0.00\% & 25.43\% & 25.43\% & 14.55\% & 26.67\% & 0.00\% \\
                   & 3 & 1.90\% & 36.40\% & 4.62\% & 36.15\% & 37.69\% & 0.00\% & 45.08\% & 29.85\% \\
                   & 4 & 35.40\% & 41.40\% & 6.00\% & 2.00\% & 8.00\% & 56.76\% & 41.27\% & 21.62\% \\
                   & 5 & 48.90\% & 43.80\% & 0.00\% & 96.67\% & 96.67\% & 80.00\% & 0.00\% & 31.71\% \\
                   & 6 & 57.20\% & 48.60\% & 0.00\% & 0.00\% & 0.00\% & 82.98\% & 0.00\% & 0.00\% \\\cmidrule{2-10}
                   & \textit{avg} & 28.68\% & 41.06\% & 2.12\% & 32.05\% & 33.56\% & 46.86\% & 22.60\% & 16.64\% \\\hline
\multirow{6}{*}{C} & 2 & 45.80\% & 67.30\% & 4.00\% & 17.00\% & 17.00\% & 36.54\% & 5.45\% & 0.00\% \\
                   & 3 & 61.40\% & 82.50\% & 45.30\% & 16.19\% & 52.57\% & 71.70\% & 0.00\% & 3.77\% \\
                   & 4 & 68.90\% & 86.40\% & 41.78\% & 58.69\% & 70.20\% & 58.59\% & 0.00\% & 1.49\% \\
                   & 5 & 72.90\% & 84.20\% & 33.06\% & 59.52\% & 64.97\% & 58.14\% & 25.86\% & 27.91\% \\
                   & 6 & 79.90\% & 86.30\% & 51.98\% & 65.87\% & 79.89\% & 100.00\% & 0.00\% & 51.85\% \\\cmidrule{2-10}
                   & \textit{avg} & 65.78\% & 81.34\% & 35.23\% & 43.46\% & 56.93\% & 64.99\% & 6.26\% & 17.00\% \\\hline
\multirow{6}{*}{D} & 2 & 10.90\% & 67.30\% & 3.08\% & 84.03\% & 82.11\% & 85.19\% & 16.92\% & 3.70\% \\
                   & 3 & 21.70\% & 68.20\% & 38.33\% & 76.67\% & 86.67\% & 38.24\% & 0.00\% & 8.82\% \\
                   & 4 & 31.90\% & 70.60\% & 47.73\% & 63.64\% & 73.03\% & 35.09\% & 8.06\% & 3.51\% \\
                   & 5 & 38.60\% & 69.70\% & 33.56\% & 77.50\% & 83.33\% & 100.00\% & 0.00\% & 13.95\% \\
                   & 6 & 56.80\% & 73.40\% & 0.00\% & 89.61\% & 89.61\% & 95.59\% & 0.00\% & 0.00\% \\\cmidrule{2-10}
                   & avg & 31.98\% & 69.84\% & 24.54\% & 78.29\% & 82.95\% & 70.82\% & 5.00\% & 6.00\% \\\hline
\multirow{6}{*}{E} & 2 & 12.70\% & 42.20\% & 0.00\% & 0.00\% & 0.00\% & 32.26\% & 0.00\% & 0.00\% \\
                   & 3 & 47.90\% & 45.90\% & 0.00\% & 0.00\% & 0.00\% & 7.50\% & 43.66\% & 20.00\% \\
                   & 4 & 58.10\% & 46.50\% & 2.33\% & 88.89\% & 88.89\% & 85.92\% & 28.28\% & 43.66\% \\
                   & 5 & 62.10\% & 80.50\% & 2.27\% & 70.00\% & 70.00\% & 54.55\% & 0.00\% & 45.45\% \\
                   & 6 & 55.20\% & 79.20\% & 3.70\% & 71.43\% & 73.28\% & 75.76\% & 0.00\% & 0.00\% \\\cmidrule{2-10}
                   & \textit{avg} & 47.20\% & 58.86\% & 1.66\% & 46.06\% & 46.43\% & 51.20\% & 14.39\% & 21.82\% \\\hline
\multirow{6}{*}{F} & 2 & 32.70\% & 69.00\% & 26.00\% & 0.00\% & 26.00\% & 27.78\% & 0.00\% & 0.00\% \\
                   & 3 & 36.70\% & 74.30\% & 25.37\% & 31.56\% & 47.14\% & 7.22\% & 0.00\% & 1.03\% \\
                   & 4 & 42.40\% & 68.20\% & 21.58\% & 22.32\% & 37.95\% & 71.95\% & 16.33\% & 2.44\% \\
                   & 5 & 55.90\% & 73.70\% & 19.76\% & 88.87\% & 91.26\% & 40.63\% & 39.62\% & 0.00\% \\
                   & 6 & 51.70\% & 74.90\% & 1.59\% & 31.75\% & 37.04\% & 100.00\% & 0.00\% & 45.83\% \\\cmidrule{2-10}
                   & \textit{avg} & 43.88\% & 72.02\% & 18.86\% & 34.90\% & 47.88\% & 49.51\% & 11.19\% & 9.86\% \\\hline
\multirow{6}{*}{G} & 2 & 57.20\% & 32.30\% & 23.33\% & 5.56\% & 28.89\% & 88.37\% & 10.42\% & 0.00\% \\
                   & 3 & 60.80\% & 31.40\% & 18.06\% & 8.33\% & 26.39\% & 100.00\% & 0.00\% & 0.00\% \\
                   & 4 & 65.70\% & 38.60\% & 0.00\% & 0.00\% & 0.00\% & 21.74\% & 36.11\% & 21.74\% \\
                   & 5 & 69.50\% & 39.80\% & 46.67\% & 0.00\% & 46.67\% & 38.24\% & 0.00\% & 0.00\% \\
                   & 6 & 74.50\% & 40.50\% & 14.81\% & 4.44\% & 19.26\% & 81.08\% & 0.00\% & 8.11\% \\\cmidrule{2-10}
                   & \textit{avg} & 65.54\% & 36.52\% & 20.57\% & 3.67\% & 24.24\% & 65.89\% & 9.31\% & 5.97\% \\\hline
\multirow{6}{*}{H} & 2 & 11.70\% & 51.50\% & 4.76\% & 0.00\% & 53.57\% & 75.86\% & 0.00\% & 0.00\% \\
                   & 3 & 30.50\% & 58.70\% & 60.00\% & 77.78\% & 92.13\% & 64.41\% & 25.32\% & 0.00\% \\
                   & 4 & 30.20\% & 47.00\% & 58.73\% & 56.13\% & 78.97\% & 76.44\% & 0.00\% & 11.49\% \\
                   & 5 & 32.20\% & 72.80\% & 62.50\% & 77.78\% & 86.67\% & 83.67\% & 0.00\% & 0.00\% \\
                   & 6 & 33.90\% & 74.60\% & 50.83\% & 98.21\% & 98.21\% & 87.69\% & 24.42\% & 0.00\% \\\cmidrule{2-10}
                   & \textit{avg} & 27.70\% & 60.92\% & 47.36\% & 61.98\% & 81.91\% & 77.61\% & 9.95\% & 2.30\% \\
\end{tabular}
\end{table}

\subsubsection{Unit testing (FEUT, BEUT)}
Teams A, C, D and F maintained a relatively high back-end testing (BEUT) coverage in average, close to or above 70\% for all sprints, with C also putting significant effort in their front-end testing (FEUT). Team B had a low test coverage overall, slowly reaching 57.2\% (FEUT) and 48.6\% (BEUT) by the end of the project. Team E and Team G had low coverage overall with reversed focus, Team E invested more in their back-end, where Team G focused more on their front-end. Team H slowly improved their FEUT over the course, still keeping a low coverage on that side of the code, but significantly improved their back-end testing for the final two sprints, with a back-end coverage above 70\%. Team G was the only team with a higher FEUT than BEUT in average. For all but two teams, the average back-end coverage is above 60\% where only two teams achieved a similar result for front-end coverage.

\subsubsection{Automated acceptance testing (AAT)}
AAT coverage values remained low for all teams. Two teams (B and E) never engaged much with automated acceptance testing overall. Three teams (A, C and H) did a better job with some sprints showing AAT coverage above 50\%, especially Team H for all but its first sprint. The remaining three teams (D, F and G) had a similar behavioural pattern with some effort put at first or sporadically, but disengaged with automated acceptance testing later on.

\subsubsection{Manual acceptance testing (MAT)}
\label{sec:mat}
Team C and H had a generally increasing manual testing coverage, with sustained effort across all sprints. Teams D and H especially increased their coverage during the last sprint with 90\% and 98\% coverage respectively, Team D maintained a high coverage, with a small drop in sprint 4 at 64\%, then increasing for the last two sprints. Teams A and E started to put effort in manual testing from sprint 4, but Team A kept a low coverage, where Team E stayed above 70\% MAT. Teams B and F had a push in manual coverage in sprint 5, but relatively low coverage elsewhere. Overall, teams' manual test coverage differed quite a lot between teams and between sprints.

\subsubsection{Overall acceptance testing (OAT)}
For all teams, we observe an overlap between automated acceptance and manual testing, sometimes significant, where teams likely tested the same acceptance criteria with both techniques. Except from Team G, we observe relatively close values between their overall coverage (OAT) and either their automated (AAT) or their manual acceptance (MAT) coverage. Teams B and F had erratic manual and overall testing behaviours where they significantly increased their acceptance test coverage in sprint 5 only. Team E never engaged with automated acceptance testing, but pushed their manual testing coverage in the last three sprints, hence a high OAT coverage for those sprints.

\subsubsection{Additional observations about testing effort}
Team D completely switched to manual acceptance testing in sprint 5, but recorded no acceptance testing for the final sprint with marginally improved unit testing coverage. Team B did not record any manual testing for sprint 6. Team G recorded very low coverage of (manual) acceptance testing overall, as well as low unit testing of their back-end, shifting most of their coverage effort on the front-end of their product. Team E increased their back-end and manual testing from sprint 4, but never really engaged with automated acceptance testing.

\subsection{Evolution of Functional Suitability}

At the end of each sprint, a marker reviewed the delivered product and calculated the percentage of story points passed, deferred and re-committed for each team, as shown in Table~\ref{tbl:team-test-metrics}.

\subsubsection{Passed Story Points (PSP)}
\begin{figure}[!ht]
    \centering
    \includegraphics[width=.9\textwidth]{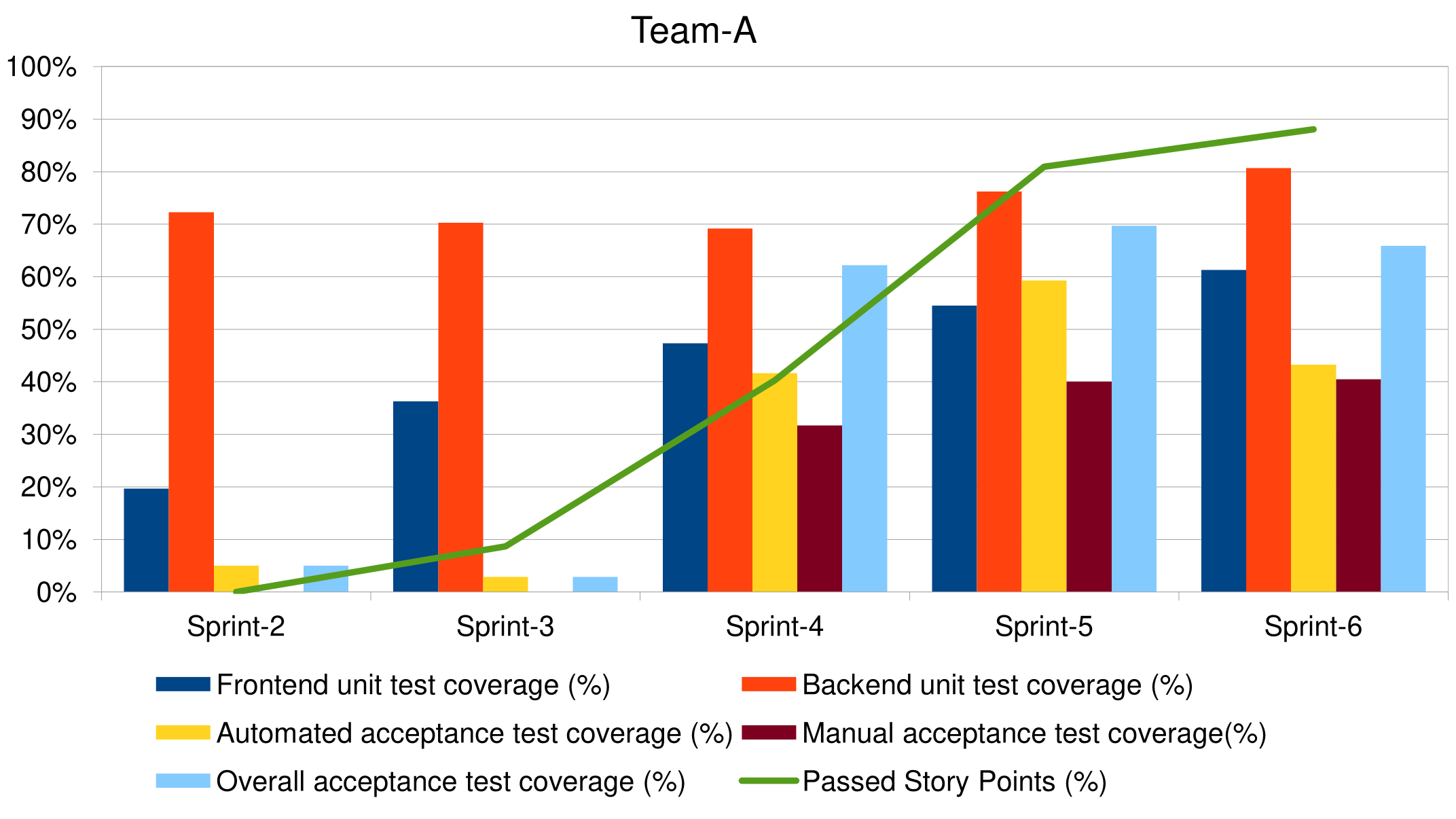}
    \caption{Evolution of testing coverage metrics and accepted stories for Team A.}
    \label{fig:team-a}
\end{figure}

\begin{figure}[!ht]
    \centering
    \includegraphics[width=.9\textwidth]{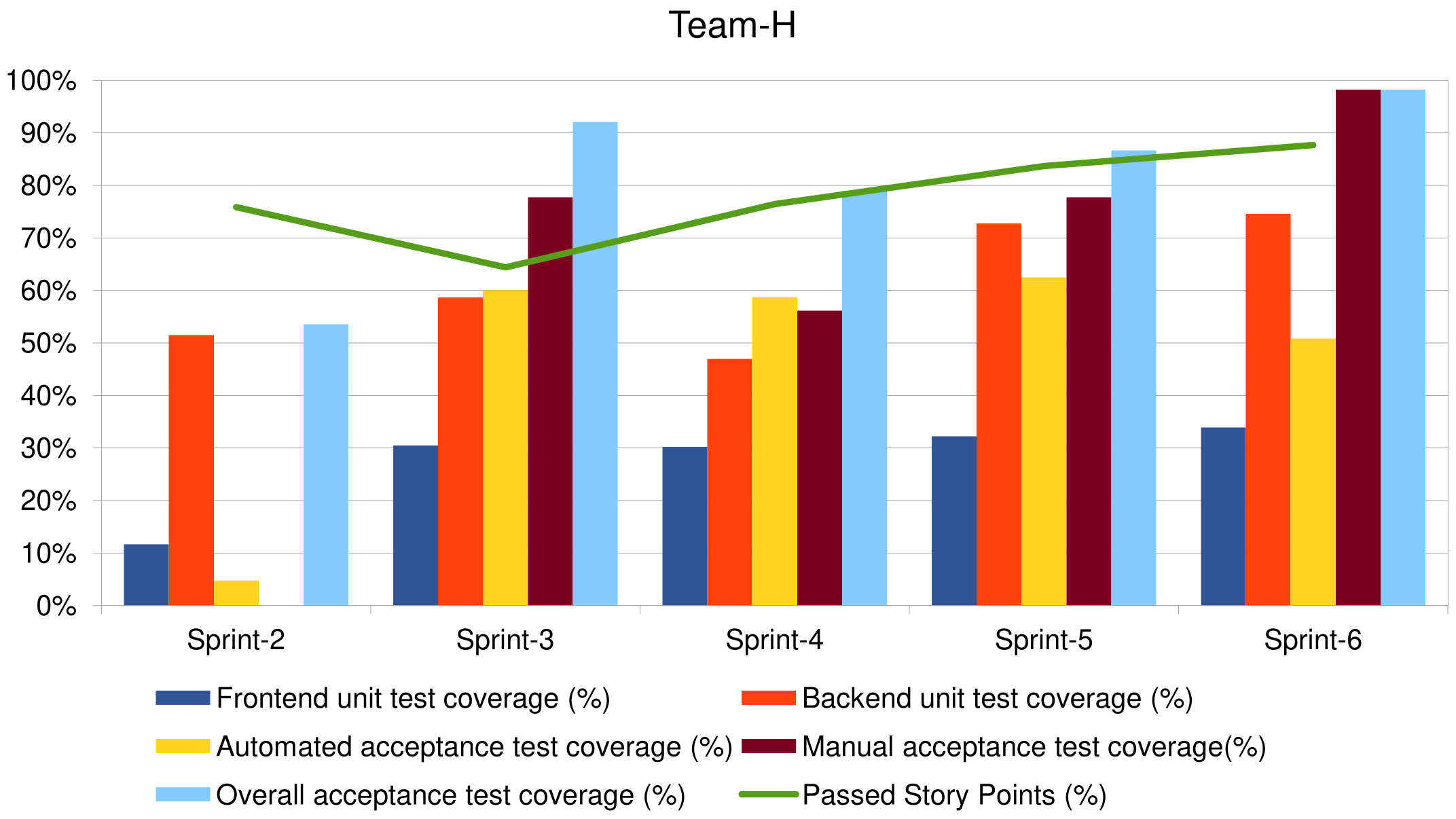}
    \caption{Evolution of testing coverage metrics and accepted stories for Team H.}
    \label{fig:team-h}
\end{figure}

Teams A and B had a slow start but steadily improved over sprints with a high percentage of stories delivered in their last sprint (88\% and 83\% resp.). Interestingly, Team A, and to a lower extent Team B, started to pass significantly more stories when they started to show evidence of better test coverage metrics (mostly AAT and OAT for Team A, FEUT for Team B), as visible in Figure~\ref{fig:team-a}. However no similar trend is visible for any other teams. Teams C, D, E, F and G were irregular in their delivered stories, Team D being the second most successful team, with high BEUT and OAT. Teams C and F started low, but passed all their stories in the last sprint, with Team C showing the highest automated testing coverage values for that sprint across all teams, where F had very low acceptance testing. Team E delivered well in Sprints 4 and 6 only. Team D and G followed a similar pattern where they both started well, then dropped significantly for two sprints (Sprints 3 and 4 for Team D, and 4 and 5 for Team G), but finished well in the last sprint, with 96\% and 81\% passed story points respectively without observable relation to their testing metrics. Team H maintained a relatively high percentage of passed story points over all sprints, with their worst delivery at 64\% in sprint 3 and an average of 77.6\% points delivered, as visible in Figure~\ref{fig:team-h}. However, they had the lowest FEUT coverage, fifth BEUT coverage, but the highest AAT (47.6\%) and OAT (72.2\%) coverage in average.

\subsubsection{Deferred Story Points (DSP)}
All teams had two deferrals, except for Team B with 3 deferrals (in sprints 2, 3 and 4) and Team A with only one in sprint 4. Teams C, F and G had a similar deferral record with two deferrals, a smaller one followed by a larger one later on. Teams D and E had also two deferrals, but their first one was larger than their second one. Team H was the only team deferring stories in the last sprint. Team B deferred an average of 22.6\%, deferring the largest amount of story points across all teams in all sprints they deferred, where Team A deferred the lowest in average (4.2\%). There is no visible trend or relation between the DSP and the test metrics from visual observation.

\subsubsection{Re-committed Story Points (RSP)}
Re-committed stories typically accounted for a lower percentage than the combination of failed and deferred points from the previous sprint for all teams. Furthermore, some teams did not explicitly recommitted to failed stories, but ``absorbed'' the fixes into other stories, i.e. they did not recommit to the whole story, but either dropped it or combined it with another related story. 

Team E is the only team where we observe a visible relation between failed or deferred stories in one sprint, and an increase of re-committed story points in the next sprint, hence the highest value for recommitted story points. Team C explicitly recommitted to a big part of their stories in the last two sprints, where Team F did the same for their last sprint only, both teams' last commitment being made of more than 45\% of re-committed stories. Apart from Team E where an increase in re-committed story points partially matched an increase in testing, no other team showed noticeable trends.

\subsection{Evolution of Time Effort}
\label{sec:evol-time-effort}

Table~\ref{tbl:team-effort-metrics} summarises the time spent by teams on the tasks of interest compared to the total time worked during that sprint (in percentages), according to students' work-logs. As mentioned in Section~\ref{sec:time-effort}, some work-logs may contain more than one tag, so we cannot sum up hours with different tags for a sprint since some recorded hours would be counted twice. However, we calculated average effort values to support our discussion in comparison to passed stories (PSP) and testing coverage values (see Table~\ref{tbl:team-test-metrics} for BEUT, FEUT, AAT, MAT).

\begin{table}[!ht]
\caption{Effort (ATE, MTE, FE, IE) and stories-related (SP, DP, RP) metrics.}
\label{tbl:team-effort-metrics}

\begin{tabular}{c r | r r r r | r r r}
Team & S\# & ATE & MTE & FE & IE & PSP & DSP & RSP \\\hline
\multirow{6}{*}{A} & 2 & 9.67\% & 11.55\% & 9.30\% & 25.75\% & 0.00\% & 0.00\% & 0.00\% \\
                  & 3 & 20.11\% & 2.12\% & 5.48\% & 23.37\% & 8.68\% & 0.00\% & 23.08\% \\
                  & 4 & 39.18\% & 2.70\% & 27.18\% & 44.59\% & 40.26\% & 21.43\% & 0.00\% \\
                  & 5 & 30.80\% & 0.18\% & 21.53\% & 37.33\% & 80.95\% & 0.00\% & 11.90\% \\
                  & 6 & 27.09\% & 1.28\% & 13.61\% & 38.87\% & 88.06\% & 0.00\% & 0.00\% \\\cmidrule{2-9}
                  & avg & 25.37\% & 3.56\% & 15.42\% & 33.98\% & 43.59\% & 4.29\% & 7.00\% \\\hline

\multirow{6}{*}{B} & 2 & 5.62\% & 1.62\% & 11.39\% & 31.99\% & 14.55\% & 26.67\% & 0.00\% \\
                  & 3 & 14.18\% & 1.20\% & 9.82\% & 18.25\% & 0.00\% & 45.08\% & 29.85\% \\
                  & 4 & 18.19\% & 2.44\% & 18.91\% & 37.79\% & 56.76\% & 41.27\% & 21.62\% \\
                  & 5 & 25.27\% & 0.14\% & 15.04\% & 33.18\% & 80.00\% & 0.00\% & 31.71\% \\
                  & 6 & 24.86\% & 0.00\% & 15.98\% & 39.18\% & 82.98\% & 0.00\% & 0.00\% \\\cmidrule{2-9}
                  & avg & 17.62\% & 1.08\% & 14.23\% & 32.08\% & 46.86\% & 22.60\% & 16.64\% \\\hline

\multirow{6}{*}{C} & 2 & 21.45\% & 1.28\% & 14.34\% & 33.66\% & 36.54\% & 5.45\% & 0.00\% \\
                  & 3 & 29.68\% & 1.07\% & 12.96\% & 29.19\% & 71.70\% & 0.00\% & 3.77\% \\
                  & 4 & 26.28\% & 2.11\% & 18.54\% & 35.46\% & 58.59\% & 0.00\% & 1.49\% \\
                  & 5 & 26.28\% & 2.93\% & 18.42\% & 35.47\% & 58.14\% & 25.86\% & 27.91\% \\
                  & 6 & 25.64\% & 2.82\% & 19.69\% & 43.25\% & 100.00\% & 0.00\% & 51.85\% \\\cmidrule{2-9}
                  & avg & 25.87\% & 2.04\% & 16.79\% & 35.41\% & 64.99\% & 6.26\% & 17.00\% \\\hline

\multirow{6}{*}{D} & 2 & 20.15\% & 17.22\% & 30.61\% & 44.92\% & 85.19\% & 16.92\% & 3.70\% \\
                  & 3 & 15.51\% & 15.47\% & 19.65\% & 30.01\% & 38.24\% & 0.00\% & 8.82\% \\
                  & 4 & 21.90\% & 3.86\% & 10.28\% & 35.94\% & 35.09\% & 8.06\% & 3.51\% \\
                  & 5 & 19.42\% & 4.30\% & 12.29\% & 35.22\% & 100.00\% & 0.00\% & 13.95\% \\
                  & 6 & 24.73\% & 4.47\% & 12.77\% & 29.90\% & 95.59\% & 0.00\% & 0.00\% \\\cmidrule{2-9}
                  & avg & 20.34\% & 9.06\% & 17.12\% & 35.19\% & 70.82\% & 5.00\% & 6.00\% \\\hline

\multirow{6}{*}{E} & 2 & 21.27\% & 1.13\% & 20.26\% & 32.82\% & 32.26\% & 0.00\% & 0.00\% \\
                  & 3 & 22.81\% & 3.11\% & 24.23\% & 55.44\% & 7.50\% & 43.66\% & 20.00\% \\
                  & 4 & 25.74\% & 3.05\% & 23.60\% & 41.72\% & 85.92\% & 28.28\% & 43.66\% \\
                  & 5 & 26.59\% & 8.02\% & 21.39\% & 49.16\% & 54.55\% & 0.00\% & 45.45\% \\
                  & 6 & 14.64\% & 8.87\% & 23.92\% & 42.09\% & 75.76\% & 0.00\% & 0.00\% \\\cmidrule{2-9}
                  & avg & 22.21\% & 4.84\% & 22.68\% & 44.25\% & 51.20\% & 14.39\% & 21.82\% \\\hline

\multirow{6}{*}{F} & 2 & 21.75\% & 2.92\% & 24.52\% & 49.00\% & 27.78\% & 0.00\% & 0.00\% \\
                  & 3 & 19.96\% & 0.00\% & 23.40\% & 47.60\% & 7.22\% & 0.00\% & 1.03\% \\
                  & 4 & 18.00\% & 1.95\% & 31.95\% & 45.27\% & 71.95\% & 16.33\% & 2.44\% \\
                  & 5 & 22.16\% & 6.95\% & 28.60\% & 47.48\% & 40.63\% & 39.62\% & 0.00\% \\
                  & 6 & 15.00\% & 0.83\% & 26.79\% & 43.96\% & 100.00\% & 0.00\% & 45.83\% \\\cmidrule{2-9}
                  & avg & 19.37\% & 2.53\% & 27.05\% & 46.66\% & 49.51\% & 11.19\% & 9.86\% \\\hline

\multirow{6}{*}{G} & 2 & 18.01\% & 0.00\% & 15.36\% & 38.63\% & 88.37\% & 10.42\% & 0.00\% \\
                  & 3 & 11.95\% & 0.00\% & 8.09\% & 29.67\% & 100.00\% & 0.00\% & 0.00\% \\
                  & 4 & 12.90\% & 0.00\% & 11.82\% & 22.23\% & 21.74\% & 36.11\% & 21.74\% \\
                  & 5 & 15.62\% & 0.00\% & 5.21\% & 20.49\% & 38.24\% & 0.00\% & 0.00\% \\
                  & 6 & 17.96\% & 0.00\% & 10.32\% & 36.04\% & 81.08\% & 0.00\% & 8.11\% \\\cmidrule{2-9}
                  & avg & 15.29\% & 0.00\% & 10.16\% & 29.41\% & 65.89\% & 9.31\% & 5.97\% \\\hline

\multirow{6}{*}{H} & 2 & 13.79\% & 3.08\% & 17.80\% & 42.64\% & 75.86\% & 0.00\% & 0.00\% \\
                  & 3 & 25.94\% & 0.98\% & 10.53\% & 28.03\% & 64.41\% & 25.32\% & 0.00\% \\
                  & 4 & 27.87\% & 3.37\% & 18.62\% & 46.11\% & 76.44\% & 0.00\% & 11.49\% \\
                  & 5 & 24.77\% & 0.80\% & 20.66\% & 33.88\% & 83.67\% & 0.00\% & 0.00\% \\
                  & 6 & 26.76\% & 3.14\% & 21.64\% & 37.87\% & 87.69\% & 24.42\% & 0.00\% \\\cmidrule{2-9}
                  & avg & 23.82\% & 2.27\% & 17.85\% & 37.71\% & 77.61\% & 9.95\% & 2.30\% \\
\end{tabular}
\end{table}

\subsubsection{Automated Testing Effort (ATE)}

Team C put a consistent effort in automated testing, with an average of 25.9\% being the highest effort across all teams, with consistently high coverage metrics (except in sprint 2). However, they had the fourth average PSP rate across all teams, at 65\%, where, on the other hand, Team G had the lowest effort at 15.3\% in average, for a slightly higher average PSP rate at 65.9\%.

Team H is another team that shows consistently high level of logged automated testing effort (except in sprint 2), with average of 23.8\%. This matches automation testing coverage data for this team for both back-end (61\% ) and automated acceptance testing (a highest overall of 47\%). They also have consistently high passing rates (PSP), with the highest average across all teams with 78\%.

Teams A and B show a similar trend in their automated testing effort (ATE) across all sprints, with a low effort at first, increasing for a few sprints, then decreasing slightly. They exhibit a higher proportion of passing story points (PSP) one sprint after putting more effort in automated testing. However, while Team B spent a slightly lower proportion of their time writing automated tests than Team A, their PSP rate was similar despite significantly higher test coverage for Team A for all of FEUT, BEUT and AAT. Team G had the lowest automated testing effort overall with Team B, below 18\%, but Team G had higher coverage values for FEUT and AAT, and the third highest value of PSP, where Team A has the second highest effort in ATE and the lowest PSP rate on average.

Teams D, E, and F had similar average ATE with comparable coverage values, except for Team E's AAT. However no observable relationship to their respective PSP can be seen, e.g., Team E had similar ATE in sprint 4 and 5 where their BEUT coverage jumped in sprint 5 with other coverage values being close to sprint 4 ones, Team F had a drop in their testing effort in sprint 6 with 15\%, but delivered all stories.

\subsubsection{Manual Testing Effort (MTE)}

Teams A and D recorded a high MTE for sprint 2, with Team D also logging more than 15\% of their time on manual testing in sprint 3. Despite consistent high MAT coverage, changes in Team D's effort does not match changes in their manual coverage. Furthermore, Team A's MAT does not match their effort either, with 0 MAT coverage in sprint 2 despite 11.5\% MTE. The other teams also have logged very little effort in manual testing, with Team G logging 0 hours during the whole project despite some (very limited) MAT coverage. Therefore, apart from Team D to some extent, we believe all other teams did not log their manual testing effort properly, so this effort data is more than likely unreliable.

\subsubsection{Fixing Effort (FE)}
\label{sec:fe}
Teams have been logging between 10.2\% (Team G) and 27.1\% (Team F) of FE in average over all sprints. Sprint 4 was generally seeing the highest proportion of FE for most teams across all sprints, even if some teams, i.e. Teams C, and G had relatively high PSP in sprint 3. We have anecdotally observed a similar trend over the past years in the course due to accumulating technical debt and teams spending time to fix UI inconsistencies. Where no observable relationship to passed stories (PSP) can be seen, it is worth noting that, from our observations, \texttt{\#fix} tags were also used for fixing technical debt by some teams, hence the numbers cannot be fully trusted.

\subsubsection{Implementation Effort (IE)}

Implementation efforts ranged from 18.3\% (Team B in Sprint 3) to 55.4\% (Team E in Sprint 3). Team G logged the lowest average implementation effort (29.4\%) where Team F logged the highest effort (46.7\%) on top of being very consistently high, above 44\% for all sprints. Five teams spent approximately a third of their time implementing features (Teams A, B, C, D, H) in average where Teams E and F logged more than 44\% IE in average. From manual observations, implementation efforts do not seem to influence or be influenced by other types of efforts, i.e. an increase in IE does not generally decrease other types of effort.


\subsection{Statistical Analysis}
\label{sec:lmer}

To help us analyse the effect of different testing metrics on functional suitability and remedial effort in subsequent sprint, we have undertaken statistical modelling. For this purpose we have used \textit{R}\footnote{Sources available next to raw data at \url{https://doi.org/10.5281/zenodo.10077456}.}, a language and environment for statistical computing. We have fitted multiple Linear Mixed-Effect models (LMER), an extension of linear regression for grouped data, as we have repeated observations of clusters of data, i.e. testing and time effort metrics per team, where teams can be considered as the random effect. Sections~\ref{sec:it-psp} and~\ref{sec:effort-psp} investigate RQ1 where Sections~\ref{sec:dsp-lmer} to~\ref{sec:psp-fix-next} relate to RQ2.

\subsubsection{Effect of Testing Metrics on Passed Story Points}
\label{sec:it-psp}

To answer RQ1, we first examine the fixed effects in terms of individual test coverage metrics on passed stories, while taking into account a random effect in terms of the observed Teams. Table~\ref{tbl:test-metrics-lmer}\footnote{The asterisks next to probability values, e.g., $**$, indicate the significance level, as output by R \texttt{summary} function.} summarises the results of the fitted model using the following function in R.

\begin{empheq}{align*}
    m \leftarrow lmer(PSP \sim FEUT+BEUT+AAT+MAT+(1|Team))
\end{empheq}

\begin{table}[htb]
\centering
\begin{tabular}{|l|l|l|l|l|l|l|}
\hline
            & Estimate & St.Error & t value & Pr(\textgreater\textpipe t\textpipe)     \\ \hline
(Intercept) & 0.3159   & 0.1723   & 1.834   & 0.07522      \\
FEUT        & 0.6323   & 0.2188   & 2.890   & 0.00657 **   \\
BEUT        & -0.3307  & 0.2950   & -1.121  & 0.26984      \\
AAT         & 0.1731   & 0.2179   & 0.794   & 0.43230      \\
MAT         & 0.3895   & 0.1401   & 2.779   & 0.00871 **   \\\hline
\end{tabular}
 \caption{Fixed effect of testing metrics on passed stories using Linear Mixed-Effect model.}
    \label{tbl:test-metrics-lmer}
\end{table}

From Table~\ref{tbl:test-metrics-lmer}, front-end unit testing (FEUT) and manual acceptance testing (MAT) have statistically significant effect on passed story points (PSP) if other involved independent variables stay unchanged. For every 1\% increase in FEUT or MAT, we can expect 0.63\% and 0.38\% increase in PSP respectively. 



The previous model investigated the effect individual testing metrics have on passed story points. We now investigate whether compound test coverage metrics can be used to predict the passed stories. We fit a model with the overall acceptance test coverage (OAT), being the intersection of automated acceptance (AAT) and manual tests (MAT), with the following function:

\begin{empheq}{align*}
    m \leftarrow lmer(PSP \sim OAT+(1|Team))
\end{empheq}

\begin{table}[htb]
\centering
\begin{tabular}{|l|l|l|l|l|l|l|}
\hline
            & Estimate & St.Error & t value & Pr(\textgreater\textpipe t\textpipe)     \\ \hline
(Intercept) & 0.34604  & 0.08367  & 4.136    & 0.000188 *** \\
OAT         & 0.46659  & 0.13721  & 3.401    & 0.001594 **  \\ \hline
\end{tabular}
\caption{Fixed effect of overall acceptance testing coverage on passed story points.}
    \label{tbl:OAT-lmer}
\end{table}

From Table~\ref{tbl:OAT-lmer}, combining the acceptance test coverage metrics together does bring some improvements in terms of predicting a successful delivery, compared to using MAT only (see Table~\ref{tbl:test-metrics-lmer}).

In order to investigate whether considering the full ``testing pyramid''\cite{Cohn2009} is a better predictor of passed story points than each coverage alone, we fit a model with the average of FEUT, BEUT and OAT testing coverage metrics, denoted AVGTC. To this end, we used the following R function, with the results summarised in Table~\ref{tbl:AVGTC-lmer}:

\begin{empheq}{align*}
    m \leftarrow lmer(PSP \sim AVGTC+(1|Team))
\end{empheq}

\begin{table}[htb]
\centering
\begin{tabular}{|l|l|l|l|l|l|l|}
\hline
            & Estimate & St.Error & t value & Pr(\textgreater\textpipe t\textpipe)     \\ \hline
(Intercept) & 0.06858  & 0.14390  & 0.477   & 0.637449     \\
AVGTC       & 0.98650  & 0.25997  & 3.795   & 0.000628 *** \\\hline
\end{tabular}
 \caption{Fixed effect of average testing coverage on passed story points.}
    \label{tbl:AVGTC-lmer}
\end{table}

The results from Table~\ref{tbl:AVGTC-lmer} show that the average test coverage has a statistically significant effect on passed stories (PSP). Moreover, the combined testing coverage has a greater effect on PSP compared to individual techniques or the compound acceptance testing (OAT), i.e. each percent increase of AVGTC would create a 0.98\% increase of PSP. In other words, in order to reliably increase the probability to pass a story, a team would need to invest in all types of testing, so cover a story at the unit, acceptance, and manual (or end-to-end) level, all together. This result is therefore in line with the commonly admitted testing pyramid concept.

\subsubsection{Effect of Testing, Fixing, and Implementation Efforts on Passed Story Points} 
\label{sec:effort-psp}

We discussed in Section~\ref{sec:evol-time-effort} that, from manual observations, time logs are potentially unreliable. However, we want to investigate this further by inspecting whether the time spent on various activities, i.e. automated testing, manual acceptance testing, fixing and implementation, has any statistically significant effect on passed story points (PSP). We applied the following R function, with its results shown in Table~\ref{tbl:effort-metrics-lmer}.

\begin{empheq}{align*}
    m \leftarrow lmer(PSP \sim ATE+MTE+FE+IE+(1|Team))
\end{empheq}

\begin{table}[htb]
\centering
\begin{tabular}{|l|l|l|l|l|l|l|}
\hline
            & Estimate & St.Error & t value & Pr(\textgreater\textpipe t\textpipe)  \\ \hline
(Intercept) & 0.1182   & 0.2662   & 0.444   & 0.660     \\
ATE         & 1.0573   & 0.8975   & 1.178   & 0.250     \\
MTE         & -0.8777  & 1.4831   & -0.592  & 0.561     \\
FE          & 0.4276   & 1.2849   & 0.333   & 0.741     \\
IE          & 0.5365   & 0.9515   & 0.564   & 0.577     \\\hline
\end{tabular}
 \caption{Fixed effect of testing, fixing, and implementation efforts on passed story points.}
    \label{tbl:effort-metrics-lmer}
\end{table}

From the results in Table~\ref{tbl:effort-metrics-lmer}, none of the recorded efforts have a statistically significant effect on past story points, i.e. $Pr\geq0.05$ for all combinations.

In a similar fashion to Section~\ref{sec:it-psp}, we aggregate the total testing effort (TTE) as the sum of automated and manual test efforts (i.e. ATE + MTE) and investigate its effect on passed story points, with the following R formula, generating the results in Table~\ref{tbl:tte-psp-lmer}. 

\begin{empheq}{align*}
    m \leftarrow lmer(PSP \sim TTE+(1|Team))
\end{empheq}

\begin{table}[htb]
\centering
\begin{tabular}{|l|l|l|l|l|l|l|}
\hline
            & Estimate & St.Error & t value & Pr(\textgreater\textpipe t\textpipe)  \\ \hline
(Intercept) & 0.3535   & 0.1779   & 1.987   & 0.0542    \\
TTE         & 0.9611   & 0.7005   & 1.372   & 0.1781    \\\hline
\end{tabular}
 \caption{Fixed effect of total testing effort on passed story points.}
    \label{tbl:tte-psp-lmer}
\end{table}

We do not see any statistically significant correlation between total testing effort and passed story points, i.e. the time students spend at writing automated tests does not predict the number of story points they will deliver.

\subsubsection{Estimation of Testing Metrics and Time Efforts on Deferred Story Points} 
\label{sec:dsp-lmer}

Our second research question looks at predicting the likeliness to under-deliver or rework stories later. Therefore, we fitted models to predict the effect of testing coverage metrics to deferred stories with the following function:

\begin{empheq}{align*}
    m \leftarrow lmer(DSP \sim FEUT+BEUT+AAT+MAT+(1|Team))
\end{empheq}

\begin{table}[htb]
\centering
\begin{tabular}{|l|l|l|l|l|l|l|}
\hline
            & Estimate  &  Std. Error & t value & Pr(\textgreater\textpipe t\textpipe)  \\ \hline  
(Intercept) & 0.29869   &  0.09353    & 3.194   & 0.00297 ** \\
FEUT        & -0.05632  &  0.11876    & -0.474  & 0.63830   \\
BEUT        & -0.30846  &  0.16013    & -1.926  & 0.06222 . \\
AAT         & -0.08233  &  0.11827    & -0.696  & 0.49094   \\
MAT         &  0.09712  &  0.07608    &  1.277  & 0.21018   \\\hline
\end{tabular}
 \caption{Fixed effect of testing metrics on deferred story points.}
    \label{tbl:it-dsp-lmer}
\end{table}

Interestingly, we see in Table~\ref{tbl:it-dsp-lmer} that back-end testing has a reversed correlation (almost significant) to deferred story points, indicating that an increase in BEUT would decrease the number of DSP. This can be interpreted as students assessing their ability to deliver all stories by looking at their test coverage preemptively. We carried similar modelling with compound testing metrics with no interesting correlation\footnote{Additional models available in the reproduction package \url{https://doi.org/10.5281/zenodo.10077456}}.

Regarding the effect of testing effort, we fitted the following model:

\begin{empheq}{align*}
    m \leftarrow lmer(DSP \sim ATE + MTE + FE + IE+(1|Team))
\end{empheq}

\begin{table}[htb]
\centering
\begin{tabular}{|l|l|l|l|l|l|l|}
\hline
            & Estimate  &  Std. Error & t value & Pr(\textgreater\textpipe t\textpipe)  \\ \hline  
(Intercept) &  0.1926   &  0.1258     & 1.531   & 0.1349  \\
ATE         & -0.3297   &  0.4125     & -0.799  & 0.4295  \\
MTE         & -0.5232   &  0.6666     & -0.785  & 0.4378  \\
FE          &  1.1141   &  0.6202     & 1.796   & 0.0811 .\\
IE          & -0.5403   &  0.4746     & -1.138  & 0.2627  \\\hline
\end{tabular}
 \caption{Fixed effect of testing efforts on deferred story points.}
    \label{tbl:effort-dsp-lmer}
\end{table}

Table~\ref{tbl:effort-dsp-lmer} suggests a close-to-significant effect between fixing effort and deferred story points. Intuitively, the more time a team spent fixing broken code, the more likely they needed to defer stories. However, as mentioned in Section~\ref{sec:time-effort} and further discussed in Section~\ref{sec:psp-fix-next}, we believe the time efforts recorded are mostly unreliable.

\subsubsection{Estimation of Testing Coverage metrics on Fixing Effort in the Following Sprint} 
\label{sec:test-fix-next}

We also want to investigate whether we can estimate the remedial effort in a subsequent sprint from testing coverage metrics and testing effort in the previous sprint. To this end, we examine the relation between the fixing effort of the next sprint (denoted $fix.next$ below) and the testing coverage metrics, i.e. BEUT, FEUT, AAT, MAT, in the current sprint with the following R formula:

\begin{empheq}{align*}
    m \leftarrow lmer(Fix.next \sim FEUT+BEUT+AAT+MAT+(1|Team))
\end{empheq}

\begin{table}[htb]
\centering
\begin{tabular}{|l|l|l|l|l|l|l|}
\hline
             & Estimate  & Std. Error & t value &Pr(\textgreater\textpipe t\textpipe)  \\\hline
(Intercept)  &  0.12414  & 0.05980    &  2.076  & 0.0593    \\
FEUT         &  0.06811  & 0.12304    &  0.554  & 0.9843    \\
BEUT         &  0.14009  & 0.13364    &  1.048  & 0.4436    \\
AAT          &  0.03481  & 0.06552    &  0.531  & 0.8133    \\
MAT          &  0.07238  & 0.07941    &  0.911  & 0.5240    \\ \hline
\end{tabular}
 \caption{Fixed effect of testing metrics on fixing effort in the following sprint.}
    \label{tbl:test-fix-next-lmer}
\end{table}

The results of the LMER model are presented in the Table \ref{tbl:test-fix-next-lmer}. None of the testing metrics have a statistically significant effect on the fixing effort in the following sprint, following our intuition from the analysis in Section~\ref{sec:fe}. A similar fitting for OAT and AVGTC (not reproduced) do not show any statistically significant results either. However, at this stage, it remains unclear whether this is due to poor logging and tagging from students.

Additionally, we investigate whether the total testing effort (TTE) influences the fixing effort in the following sprint. We created the following model in R with its result shown in Table~\ref{tbl:tte-fix-next-lmer}.

\begin{empheq}{align*}
    m \leftarrow lmer(Fix.next \sim TTE+(1|Team)) 
\end{empheq}

\begin{table}[htb]
\centering
\begin{tabular}{|l|l|l|l|l|l|l|}
\hline
            & Estimate & St.Error & t value & Pr(\textgreater\textpipe t\textpipe)  \\ \hline
(Intercept) & 0.10093  & 0.03877  & 2.603   & 0.0147 *  \\
TTE         & 0.30828  & 0.13904  & 2.217   & 0.0346 *  \\\hline
\end{tabular}
\caption{Fixed effect of total testing effort on fixing effort in the following sprint using Linear Mixed-Effect model}
    \label{tbl:tte-fix-next-lmer}
\end{table}

\begin{figure}[!ht]
\centering
\includegraphics[width=.8\textwidth]{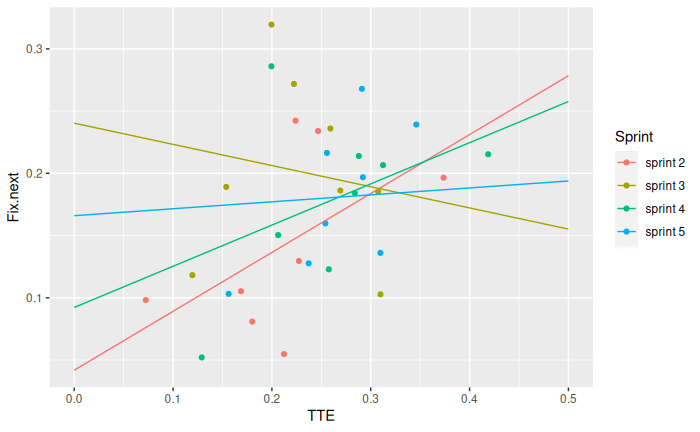}
\caption{Effect of Total Testing Effort (TTE) on next sprint's fixing effort (Fix.next), per sprint.}
\label{fig:TTE_on_Fix_per_sprint}
\end{figure}

From Table~\ref{tbl:tte-fix-next-lmer} we see that an increase in overall testing effort in one sprint predicts a slight increase in fixing effort during the next sprint. To further inspect the relationship between testing and fixing effort, we also investigate the same trend per individual sprint, as visible in Figure~\ref{fig:TTE_on_Fix_per_sprint}, where the coloured lines represents the trend for all teams per sprint.

As we can see from Figure~\ref{fig:TTE_on_Fix_per_sprint}, for all but Sprint 3, the fixing effort in the next sprint (denoted ``Fix.next'') increases as the total testing effort (TTE) increases. This may further support our initial intuition regarding incorrect and unreliable effort logging from students. Another plausible explanation would be that tests must be rewritten or adapted the next sprint, so an increase in automated tests increases the test code base that needs to be maintained later on. 

\subsubsection{Estimation of fixing effort in the following sprint based on passed story points in the current sprint}
\label{sec:psp-fix-next}

Following up with the potential conflicting results regarding testing efforts depicted in Sections~\ref{sec:dsp-lmer} and~\ref{sec:test-fix-next}, we investigate whether the amount of code fixing, i.e. remedial, work in the following sprint is influenced by the number of story points a team passed the previous sprint. The goal of this analysis is to further verify our intuition around bad time logging and tagging. In other words, intuitively, we would expect a higher proportion of time spent implementing fixes after a failed sprint delivery where a successful delivery would mean less time spent on fixes the next sprint. To this end, we run the following R function, with its results shown in Table~\ref{tbl:psp-fix-next-lmer}.

\begin{empheq}{align*}
    m \leftarrow lmer(Fix.next \sim PSP+(1|Team))
\end{empheq}

\begin{table}[htb]
\centering
\begin{tabular}{|l|l|l|l|l|l|l|}
\hline
            & Estimate & St.Error & t value & Pr(\textgreater\textpipe t\textpipe)      \\ \hline
(Intercept) & 0.17056  & 0.02624  & 6.500   & 9.09e-06 ***  \\
PSP         & 0.01043  & 0.03139  & 0.332   & 0.742         \\\hline
\end{tabular}
 \caption{Fixed effect of passed story points on next sprint's fixing effort.}
    \label{tbl:psp-fix-next-lmer}
\end{table}

Similarly to Table~\ref{tbl:effort-metrics-lmer} where the time effort logged by students does not show statistically significant correlation to functional suitability, passed stories in one sprint does not statistically relate to the fixing effort in the following sprint. We believe that this lack of correlation further support our observation regarding inaccurate logging and tagging of work-logs by students.

\section{Discussion}
\label{sec:discussion}

We first combine the findings from our manual observations and statistical modelling in relation to our initial research questions (see Section~\ref{sec:rq}), then discuss the limitations of our empirical study, as well as the threats to validity, following Wohlin et al's guidelines~\cite{Wohlin2012}.

\subsection{Summary of Findings}

From the existing literature, as discussed in Section~\ref{sec:rw}, we know that Test- (TDD) and Behaviour-Driven Development (BDD) have shown positive impacts on external product quality in the industry~\cite{Dookhun2019,Farooq2023}. However, similar studies in an academic context reported mixed results when applying TDD, where only Rocha et al. reported positive effects to students' grades~\cite{Rocha2021}. In this research, we investigated whether we could observe beneficial impacts of both software development techniques to the software products created in a year-long project designed to replicate an industry-level Scrum-based project. 

\subsubsection{RQ1 - Combination of Testing Metrics to Predict Functional Suitability}

When combining the manual and statistical analysis, we observed that a high level of overall acceptance testing (OAT), calculated as a combination of manual and automated acceptance testing, is present for some teams that delivered a higher quality product. This was particularly the case for 3 out of 8 teams, i.e. Teams A, and to a lower extent E, where higher passing rate was observed at the same time as their overall acceptance testing coverage went up, and Team H maintained a high passing rate with a consistent high overall acceptance testing coverage. However, Team D also had a high OAT across sprints, but irregular deliverable passing rate.

When analysing individual testing metrics, front-end and manual testing are statistically significant according to our models, where back-end testing has a negative impact to passed stories. 

\begin{tcolorbox}
    This suggests that testing techniques that are close to user experience seem to be more helpful in achieving high level of story acceptance rate in the early stages of project development for our population of junior developers in an academic team-based project.
\end{tcolorbox}

On the other hand, any combinations of testing coverage with low or inconsistent acceptance testing would be combined with low quality products (e.g, Teams D, F). 
\begin{tcolorbox}
    This suggests that teams of junior developers have to regularly and incrementally acceptance test their software products. This testing needs to be performed from the early stages of the product development to unsure a satisfactory product with functional suitability and stable deliveries from sprints to sprints.
\end{tcolorbox}

\subsubsection{RQ2 - Relation between Testing and Remedial Effort}

From both manual and statistical analysis, we cannot draw any strong correlations between students' work-logs and functional suitability (i.e. passed story points), or remedial effort in the next sprint. We observed that some teams disengaged with automated acceptance testing after few sprints (e.g., Teams D, F), shifting their focus to manual (acceptance) testing, with some noticeable improvements for Team D. However, increases in manual testing coverage was typically disconnected from the logged manual testing effort. Furthermore, the automated testing effort was recorded all together (i.e. unit and acceptance testing), where variations in coverage seemed disconnected from recorded time effort. This assumption is further substantiated by our analysis of the effect of failed deliveries, i.e. low passed stories, to the remedial effort, i.e. amount of \texttt{\#fix} effort in the next sprint. We assume that students work-logs are unreliable in their current form for the following two reasons: (i) students did not record their time appropriately, either because of wrong application of expected work-log tags, e.g., usage of \texttt{\#test} where \texttt{\#testmanual} would apply, or bad logging hygiene, (ii) students gamifying their work-logs to meet the course requirements in terms of practising all aspects of software development. Still, we identified a potential effect between decreasing back-end test coverage and an increasing probability to defer stories.

\begin{tcolorbox}
    We cannot identify any significant relations between testing coverage metrics and time effort logged on one hand, and neither of functional suitability or remedial effort on the other hand with the data we gathered in the context of a year-long software development project course. However, our data suggests that back-end testing coverage is a potential indicator of stories deferrals where an increase in back-end testing lowers the probability to defer stories. 
\end{tcolorbox}

\subsection{Implications and Recommendations}

In this section, we discuss additional implications relevant to educators and software engineering practitioners that this study suggests.

\subsubsection{Implication to Educators}

Behaviour-Driven Development (BDD) is a complex practice, by nature. Despite the technique being introduced in the co-requisite course with dedicated assessment items, students struggle to apply BDD to a more complex scenario. The low overall engagement across teams with automated acceptance testing coupled with gradual shift to manual testing after a few sprints demonstrate that more training is required to fully engage as well as potentially attain beneficial effects from this technique. Additionally, even if the project runs for a full academic year, the size of the products by the end of the year are still relatively small compared to real-world software (approx 20 to 30K LOC). Therefore, manual testing remained a viable and, perceivably faster testing method from both the students' point of view, and the statistical analysis we conducted. This observation has been anecdotally confirmed by students during end-of-year interviews. Still, teams that delayed their testing effort tended to take more time to achieve a higher functional suitability, but eventually caught up if their testing metrics improved. This is in line with Kazerouni et al's finding where the authors suggested that earlier and regular testing leads to higher product quality~\cite{Kazerouni2017}.

\begin{tcolorbox}
    We recommend educators to encourage students to write automated tests early on to maintain a higher level of functional suitability with minimum effort. 
\end{tcolorbox}

Despite the relatively long period with a sizeable software product, we believe that students did not have enough time and opportunity to learn how to properly implement automated acceptance testing in more complex scenarios where BDD is implemented. This can be linked to the students' shared inexperience with that particular practice preventing them to reach the benefits highlighted in industry contexts identified by Farooq et al~\cite{Farooq2023}. 

\begin{tcolorbox}
    We recommend educators interested in teaching BDD and automated acceptance testing to limit, but not remove, students' ability to manually test their software to prevent an ``easy'' alternative to automated acceptance testing.
\end{tcolorbox}

\subsubsection{Implications to Practice}

To a larger extent to Dookhun et al.'s observed decrease in productivity when applying BDD~\cite{Dookhun2019}, students in the course showed a disengagement regarding automated acceptance testing as their software was growing, led by a perceived burden and lack of knowledge regarding automated acceptance testing. However, our data show that both unit front-end and manual acceptance test coverage are good predictors of a higher functional suitability.

\begin{tcolorbox}
    Similar to prior studies, our results show that tests that are close to the users' perspective, including unit front-end testing, are beneficial to a high functional suitability, i.e. external quality of software. 
\end{tcolorbox}

We also observed that there is ``no metric to rule them all'', i.e. no single test coverage or work effort metric can strongly predict the success of a software product delivery in terms of its functional suitability. Where front-end and manual testing showed some statistical correlation, the best prediction resulted from the combination of all coverage metrics, with almost a 1:1 ratio, i.e.

\begin{tcolorbox}
    In order to achieve a high functional suitability, user stories (or features) must be covered by tests at the unit, acceptance and manual-level, covering the full testing pyramid.
\end{tcolorbox}

\subsection{Limitations}
\label{sec:limitations}

Current work exposes some limitations to be considered when assessing our findings and recommendations. There are also threats to validity, some of which we have mitigated with counter-measures, as we discuss in this section.

\subsubsection{Limitations to Findings}
\label{sec:limitations-findings}
\paragraph{Manually Calculated Acceptance Testing Metrics} 

Acceptance test coverage records for both manual and automated acceptance testing were determined by manually mapping written test descriptions to acceptance criteria. Some tests were clearly numbered according to their corresponding criteria, others not, so some manual judgement was necessary at times. Acceptance criteria were considered fully covered if there was at least one manual or automated acceptance scenario written for it. However, the quality of acceptance tests was not assessed such that some tests may cover blue-sky scenarios only, hence coverage metrics are potentially optimistic. 

\paragraph{Students' Work-log} 

Work-logs can hold multiple tags, so the time effort data we gathered has limitations by construct. Furthermore, as mentioned in Section~\ref{sec:dev-process}, students are expected to engage with all aspects of software development, which is part of their grading in the project course. Some students may have tried to gamify the system by logging hours based on our expectations rather than on their actual work. However, the teaching team carried sanity checks when assessing the students performance to prevent obvious attempts to cheat.

\paragraph{Misunderstanding of Tags}

We observed Teams with high manual testing coverage recording only very little to no manual testing effort. Other technical tags such as \texttt{\#fix} or \texttt{\#chore} (not used in this study) were also often misused by students, which has had an influence on the time effort data, as discussed previously.

\paragraph{Rotation of Product Markers} 

End-of-sprint products were assessed by different markers, which were generally rotating between teams and sprints. However, the rotation was not following a strict round-robin. Despite a common set of acceptance criteria shared by all markers, as well as a coordinated effort when assessing borderline cases, variation in assessments may have occurred, which may have influenced the values calculated for the passed story points (PSP) metric. 

\subsubsection{Threats to Validity}

\paragraph{Conclusion Validity}

Conclusion validity refers to the ability of the experimenter to draw conclusions from the evidence gathered.

\begin{description}
    \item[Statistics:] We combined both manual observations and statistical models to draw conclusions regarding our research questions. Still, we don't claim statistical power as our population is too small. We also avoided to discuss the statistical results from data (i.e. time efforts) that was visibly unreliable.
    
    \item[Reliability of treatment:] All teams received the same teaching, had access to the same resources, and worked on a similar product with the same stakeholders. We also applied a rotation of markers across sprints which partially mitigates divergences in the assessment of the delivered software product.
\end{description}

\paragraph{Internal Validity}

Internal validity refers to influences to conclusions (i.e. causality) drawn from the evidence gathered.

\begin{description}
\item [Instrumentation:] As mentioned in Section~\ref{sec:limitations-findings}, we manually mapped tests to acceptance criteria and assumed that an AC was covered when at least one manual or automated acceptance test was written for it. This process was carried by the first author, with random cross-checks by the second author. All other data (i.e. unit coverage and work-logs) were calculated automatically from professional tools (i.e. SonarQube) or extracted from Jira, which limit human errors. 

\item [Maturation:] The project course lasted for a full academic year, so it is expected that the students' ability to test their code improves. Boredom is also generally observed around mid-year (i.e. sprint 4), where the external quality of the product generally decreases for all teams. We mitigated this threat by taking data measures at different points in time, when the students' workload in other courses was lower. We also observed multiple teams.

\item [Attrition (Mortality):] Some teams were affected more than others by drop-outs. One team ended up with four members where other teams stayed intact until the end of the project. While drop-outs in one particular sprint may have punctually affected the ability to deliver what was promised, each team had an opportunity to adjust their team's commitment, i.e. the amount of features to deliver each sprint.

\item [Social:] The course's requirements state that students have to touch to all aspects of the project. This may have led students to ``distord'' some of their work-logs, combining tags for better marks. To partially mitigate that issue, the teaching team carried sanity checks of work-logs while assessing students' performance. As discussed in Sections~\ref{sec:effort-psp} and following, we limited our conclusions regarding time efforts to strong trends we could observe from multiple perspectives.

\end{description}

\paragraph{Construct Validity}

Threats to construct validity concern (i) the ability of the experiment's design to reflect the concept under study, and (ii) subjects and experimenters' behaviours within the experiment.

\begin{description}
\item[Design:] We combined multiple metrics and observation methods to reach the conclusions and recommendations above. We also employed multiple assessors in sustained communication when evaluating the product deliveries. Additionally, due to the length of the project course, we cannot ensure that external factors (e.g., other courses' deadline) have not influenced the measures. However, the product deliveries were scattered along the academic year, so that the effects of potential interactions with external factors at specific times are partially mitigated. 

\item[Social:] Students are assessed on their ability to work on all aspects of software development which may have had an influence on their behaviour. On the other hand, students were not aware of the specific aspects of marking beyond the requirements of the course, so that they could not specifically skew these particular measures related to testing.

\end{description}

\paragraph{External Validity}

External validity relates to the ability to generalise the conclusions of the experiment.

\begin{description}

\item[Selection and treatment:] Our study targets junior software developers, and our population was composed of last year undergrad students, most of them with practical work experience through internships. However, because of the relative homogeneity of the population, we cannot generalise beyond similar populations or team settings. We also compared our conclusions to the existing body of knowledge to further substantiate our findings and recommendations.

\item[Setting and treatment:] All teams were exposed to industry-level practices, commonly used in agile software development settings. They were provided with industry-relevant tools together with training material and support from senior teaching staff knowledgeable in those techniques. We reproduced a realistic environment, but thoroughly considered and discussed its limitations and assumptions.

\end{description}

\section{Conclusion and Future work}
\label{sec:conclusion}

In this paper, we presented the results of a case study aimed at assessing which combination of testing techniques and metrics contributes to the external quality (i.e. the functional suitability) of a software product in a year long university project. We also looked at the effects between testing metrics and time efforts in one development iteration, and the subsequent effort spent fixing issues in the following iteration (sprint). The case study involved 8 teams made out of 4 to 8 students over a period of 5 sprints spanning over a full academic year. 

The results show that there is statistically significant correlations between passed story points (used as a proxy to functional suitability) and the manual acceptance and front-end unit test coverage in the same sprint. Additionally, the average testing coverage, defined as the combination of unit, automated acceptance and manual acceptance testing, is a good predictor of a successful product delivery, following the commonly admitted ``testing pyramid'' pattern. 

We observed that automated acceptance testing was not used to its full potential, with teams partially disengaging mid-year. We concluded that students were not very proficient in using this testing technique and have struggled to use it in more complex scenarios. We therefore recommended to educators interested in teaching behavioural-driven development practices to set up learning environment where students cannot drift to manual testing too easily.

However, our ability to investigate the relationship between time effort and deliveries was hindered by unreliable work-logs from students, as well as apparently misunderstood (or too complex) work-log tagging convention by students.

For future work, we plan to improve how we teach BDD and ATDD. In order to make the learning more approachable to students, we will offer more workshops on this topic at the different learning milestones to overcome typical learning obstacles around BDD and ATDD. In this research we only looked at code coverage (unit tests) and acceptance test coverage (manual testing and automated acceptance testing). In order to know how well tests cover a feature we need to look at the quality of tests too. Last, we need to improve our guidelines regarding work-logs and their tagging, with more stringent sanity checks and, were applicable, the ability to reclassify some logs, in order to replicate this study.

\section*{Data Availability Statement}

The datasets with code metrics, statistics models, and analysis used in the current study are available in the Zenodo repository, \url{https://doi.org/10.5281/zenodo.10077456}. Access to source code and worklogs are kept private for confidentiality and privacy reasons, but can be requested from the authors.

\section*{Conflict of Interest}

The authors declared that they have no conflict of interest.

\bibliographystyle{spbasic}
\bibliography{references}

\end{document}